# Optimum Design of GaAs/AlGaAs Surface-Relief VCSELs with Single-Mode Operation at 808 nm

Hassan Hooshdar Rostami
Dept. of Electrical and Computer Eng.
*Tarbiat Modares University*
Tehran, Iran
h_hooshdarrostami@modares.ac.ir

Vahid Ahmadi
Dept. of Electrical and Computer Eng.
*Tarbiat Modares University*
Tehran, Iran
v_ahmadi@modares.ac.ir

Saeed Pahlavan
Dept. of Electrical and Computer Eng.
*Tarbiat Modares University*
Tehran, Iran
s_pahlavan@modares.ac.ir

*Abstract*—This paper reports an opto-electro-thermal analysis of VCSEL structures with a surface-relief filter and different aperture diameters. 808-nm GaAs/AlGaAs VCSELs with different aperture diameters and different surface-relief filter diameters are studied and the LI curves, the optical spectra, and the temperature peaks of the devices are obtained. Also, the impact of utilizing different oxide apertures on the temperature peak is shown. The effect of the surface-relief filter diameter on VCSEL's performance is also investigated and the optimal filter diameter is obtained for maintaining single-mode operation while producing as much output power as possible.

*Keywords—Mode filter, Single-mode operation, Surface relief, VCSELs*

## I. Introduction

Vertical-cavity surface-emitting lasers (VCSELs) have emerged as key light sources for various applications, including optical interconnects, atomic clocks, and optical mice [1]. Their low threshold currents, circular output beams, and wafer-scale integration provide considerable advantages over edge-emitting lasers [2]. VCSELs have a single longitudinal mode due to their short cavity lengths but higher transverse modes appear with increasing active region volumes to reach higher output powers [3].

Various techniques have been explored to obtain fundamental mode operation in VCSELs. A seminal advancement was incorporating oxide confinement layers formed by selective lateral oxidation of AlGaAs [4]. The resulting oxide aperture enables optical and current confinement to improve laser performance and also restrict higher-order modes. Alternative confinement strategies have also been proposed using surface relief [5], Tunnel-Junction-based VCSELs [6], ion implantation [7], and high contrast gratings (HCGs) [8]. HCG VCSELs have achieved single-mode operation by selectively introducing the optical loss for higher-order modes through the grating design. Monolithic high contrast grating (MHCG) VCSELs improve confinement and performance, and they are also easier to fabricate than HCG VCSELs [9].

More recent work has explored metasurface-based VCSELs [10] and photonic crystal VCSELs [11] to obtain single fundamental mode lasing. VCSELs with fundamentally stable single-mode emissions are an active ongoing research frontier and further advances in performance, processing simplicity, and integration are desired.

In this paper, intending to design and investigate VCSELs' single mode operation at 808 nm, we perform opto-electro-thermal analysis of the structures and study the effect of VCSEL's aperture and the surface relief diameter on its performance and mode distribution. Due to their mutual interactions, the optical, thermal, and electrical analyses should be performed self-consistently. The next section of the paper describes the structure of the simulated laser. The governing equations and the VCSEL's structure are presented in the second and third sections, respectively and the fourth section is dedicated to the simulation results and discussions. The last section concludes the paper.

## II. Governing Equations

In this study, we have conducted an opto-electro-thermal analysis of the VCSEL structure. Shockley-Read-Hall recombination (SRH), Auger recombination, self-heating, and interface recombination were considered. The optical modes are determined from the solution of the scalar Helmholtz equation.

$$\left(\frac{\partial^2}{\partial x^2} + \frac{\partial^2}{\partial y^2} = k_0^2 \big(\varepsilon(x,y) - n_{eff,m}^2\big)\right) E_m(x,y,z) = 0 \quad (1)$$

where $k_0$ is the wave vector, $\varepsilon$ is the permittivity, $n_{eff,m}$ is the effective refractive index of each mode, and $E_m$ is the electric field intensity of each mode. Optical field propagation and electro-thermal transport are coupled via rate equations for the intensity within each cavity mode:

$$\frac{\partial S_{m,\omega}}{\partial t} = \left(G_{m,\omega} - \frac{1}{\tau_{w,\omega}}\right) S_{m,\omega} + R_{m,\omega}^{spon} \quad (2)$$

where the spontaneous emission into the mode is given by

$$R_{m,\omega}^{spon} = \int dV |E_m|^2 u(\omega) \quad (3)$$

and the modal gain is defined as:

$$G_{m,\omega} = \int dV |E_m|^2 \frac{c}{n_{eff,m}} g(\omega) \quad (4)$$

Here, $g$ is the material gain and $u$ is the spontaneous emission spectrum. The losses in the photon rate equation are the sum of contributions due to light leaving the cavity through the facets, light scattered out of the cavity, and absorptive losses:

$$\frac{1}{\tau_{w,\omega}} = \frac{1}{\tau_{mirror}} + \frac{1}{\tau_{scatter}} + \frac{c}{n_{eff,m}} \alpha_b \quad (5)$$

where $\alpha_b$ is the free-carrier absorption coefficient. Poisson's equation determines the electric potential $\Phi$

$$\nabla \varepsilon \nabla \phi = q(n_e + n_h - N_D^+ + N_A^-) \quad (6)$$

The non-radiative recombination $R^{dark}$ occurs in both bulk and quantum well regions. The rate is the sum of contributions from Auger and Shockley-Read-Hall (SRH) processes. In the Auger process, an electron and hole recombine while the corresponding energy is transferred to an additional electron or hole. The Auger recombination is modeled by:

$$R^{Auger} = \left(C_e^{Auger} n_e + C_h^{Auger} n_h\right)(n_e n_h - n_i^2) \quad (7)$$

where $n_i$ is the intrinsic carrier density, and $C_{e/h}^{Auger}$ are Auger coefficients. The SRH process describes recombination via trap levels and is modeled by:

$$R^{SHR} = \frac{n_e n_h - n_i^2}{\tau_h^{SHR}(n_e + n_e^{trap}) + \tau_e^{SHR}(n_h + n_h^{trap})} \quad (8)$$

where $\alpha_b$ is the free-carrier absorption coefficient, $\tau_{h/n}^{SHR}$ are the lifetimes for electrons and holes, and $n_{e/h}^{trap}$ are the trap occupations. Similar to the SRH model, the interface recombination rate (per surface area) is given by:

$$R^{interface-SHR} = \frac{n_e n_h - n_i^2}{\frac{1}{v_h^{interface-SHR}}(n_e + n_e^{trap}) + \frac{1}{v_e^{interface-SHR}}(n_h + n_h^{trap})} \quad (9)$$

where $v_{e/h}^{interface-SHR}$ are the interface recombination velocities. Due to carrier-phonon scattering, part of the electronic energy may be transferred to the crystal lattice leading to an increase of the lattice temperature. This lattice heat flow equation can be solved to describe self-heating effects:

$$\left(C_L + \frac{3}{2} k_B (n_e + n_h)\right) \frac{\partial T}{\partial t} = \nabla(\kappa_L \nabla T - \vec{S}_e - \vec{S}_h) + H \quad (10)$$

where $C_L$ is the heat capacity of the lattice, $k_B$ is the Boltzmann constant, $T$ is temperature, $\kappa_L$ is the heat conductivity of the lattice carriers, $\vec{S}_{e/h}$ are electron and hole energy fluxes, and $H$ is the sum of different heat sources.

### III. DEVICE STRUCTURE

#### A. VCSEL design

Figure 1 a shows the proposed surface-relief VCSEL structure. The mesa diameter is 33 µm, and $Al_{0.2}Ga_{0.8}As/Al_{0.9}Ga_{0.1}As$ Bragg reflectors provide optical confinement with 37.5 and 26 pairs for lower and upper mirrors, respectively. The uppermost DBR layer is etched to form the surface-relief filter. The structure is simulated with different aperture diameters throughout the manuscript. The lasing wavelength is 808 nm. The spacer layers in the active region are 96 nm thick and the oxidation layer used to form the oxide aperture is 45 nm thick. Also, three layers of $Al_{0.2}Ga_{0.6}In_{0.2}As$ quantum wells with a thickness of 6 nm and four layers of 8 nm thick $Al_{0.4}Ga_{0.6}As$ barriers are used to provide optical gain. The parameters employed in the simulations are listed in Table I.

#### B. Surface-relief filter structure

The surface-relief filter structure is obtained by etching the top layer of the top DBR as depicted in Fig. 1.b. $d_{filter}$ is the filter diameter, and $t_{filter}$ is its thickness which is a quarter-wavelength of the propagating light. In this work, we simulated the structure for different filter diameters to achieve the optimum diameter of the surface relief.

The idea of a surface-relief filter is to introduce a different phase shift for different modes in the structure. As a result, different modes experience different modal gains, leading to the attenuation of higher-order modes. By etching the top layer of the DBR, we introduce a $\pi/2$ phase for higher modes in the structure so that they will be suppressed to achieve single-mode operation. The filter diameter strongly affects the VCSEL's performance. Therefore, we need to find the best dimensions for the introduced filter.

Figure 2 shows the LI curves for different filter diameters for a 20 µm aperture. Figure 3 illustrates the mode spectrum of these VCSELs. As shown, we still see the multimode operation for a filter diameter of 0.6×$d_{mesa}$. The reason is that the surface-relief filter still spatially covers higher modes and therefore mirror reflectivities are sufficient for them to lase. The coverage diminishes as the filter diameter shrinks until we see a single-mode operation. On the other hand, the further the filter diameter shrinks, the less the filter covers the fundamental mode leading to a reduced mirror reflectivity for the fundamental mode. Hence, the output power and efficiency decrease by further reduction of filter diameter below an optimal range. Therefore, we chose the diameter of 0.55×$d_{mesa}$ as the best filter size for this structure.

Figure 4 and Fig. 5 show the LI curves and light spectra for the surface-relieved VCSELs with an aperture diameter of 22 µm and different filter diameters. As seen, the surface-relief filter with a diameter of 0.55×$d_{mesa}$ has a better output power and operates at single mode. Figure 6 to Fig. 9 depict the LI curves and the optical spectra for the surface-relief VCSELs with an aperture diameter of 24 µm and 26 µm and different filter diameters. It is shown that the 0.50×$d_{mesa}$ and 0.60×$d_{mesa}$ wide surface-relief filters are the best options for VCSELs with apertures diameters of 24 µm and 26 µm, respectively.

TABLE I. PARAMETERS USED IN THE SIMULATIONS

| | *Material* | *Thickness (nm)* | *Doping ($cm^{-3}$)* |
|---|---|---|---|
| **Top Contact** | Au | 100 | - |
| **Surface-relief filter** | $Al_{0.9}Ga_{0.1}As$ | 65 | $3\times10^{18}$ |
| **Top DBR** | $Al_{0.2}Ga_{0.8}As/Al_{0.9}Ga_{0.1}As$ (26.5 pairs) | 56/65 | $3\times10^{18}$ |
| **Top Oxide Layer** | $Al_xO_y/Al_{0.98}Ga_{0.02}As$ | 45 | $1.5\times10^{18}$ |
| **Active Region** | $Al_{0.45}Ga_{0.55}As$ (spacer) | 96 | |
| | $Al_{0.4}Ga_{0.6}As$ (barrier) | 8 | - |
| | $Al_{0.2}Ga_{0.6}In_{0.2}As$ (QW) | 6 | |
| **Bottom DBR** | $Al_{0.2}Ga_{0.8}As/Al_{0.9}Ga_{0.1}As$ (37.5 pairs) | 56/65 | $3\times10^{18}$ |
| **Bottom Contact** | Au | 100 | - |

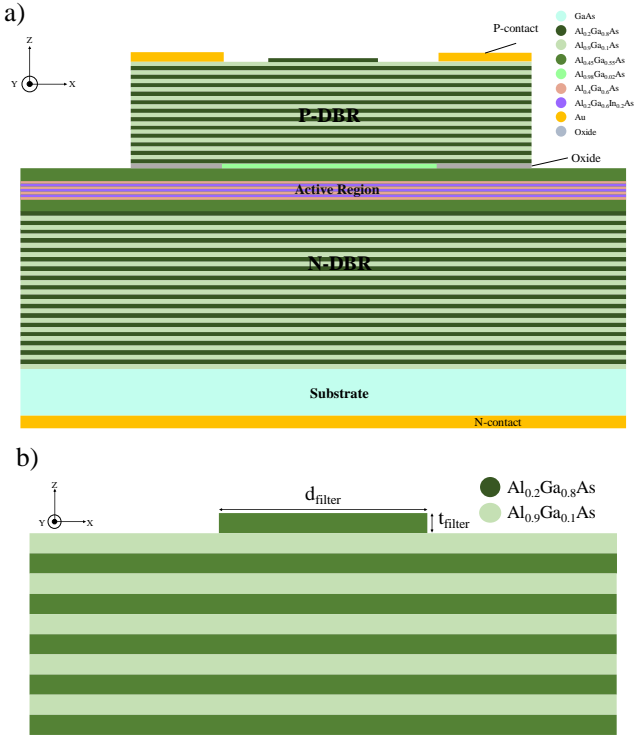

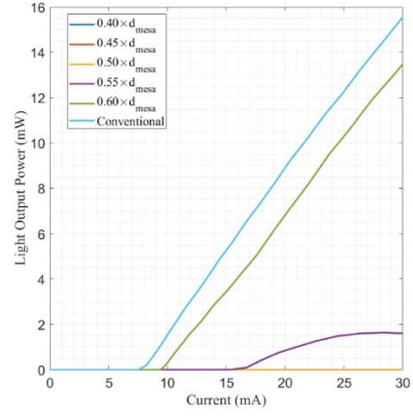

Fig. 2. LI curves for different filter diameters for a VCSEL with a 20 µm wide aperture

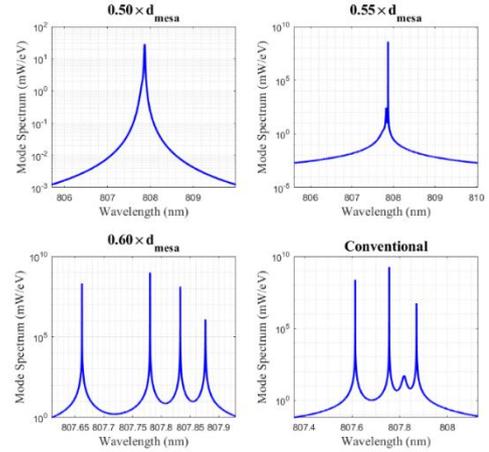

Fig. 3. Mode spectrum for different filter diameters for a VCSEL with a 20 µm wide aperture

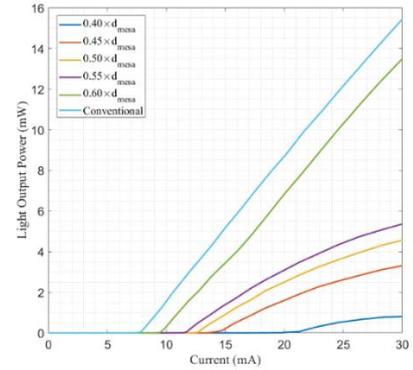

Fig. 4. LI curves for different filter diameters for a VCSEL with a 22 µm wide aperture

Fig. 1. a) Surface-relief VCSEL structure and b) surface-relief filter structure

Figure 2 shows the LI curves for different filter diameters for a 20 µm aperture. Figure 3 illustrates the mode spectrum of these VCSELs. As shown, we still see the multimode operation for a filter diameter of $0.6 \times d_{mesa}$. The reason is that the surface-relief filter still spatially covers higher modes and therefore mirror reflectivities are sufficient for them to lase. The coverage diminishes as the filter diameter shrinks until we see a single-mode operation. On the other hand, the further the filter diameter shrinks, the less the filter covers the fundamental mode leading to a reduced mirror reflectivity for the fundamental mode. Hence, the output power and efficiency decrease by further reduction of filter diameter below an optimal range. Therefore, we chose the diameter of $0.55 \times d_{mesa}$ as the best filter size for this structure.

Figure 4 and Fig. 5 show the LI curves and light spectra for the surface-relieved VCSELs with an aperture diameter of 22 µm and different filter diameters. As seen, the surface-relief filter with a diameter of $0.55 \times d_{mesa}$ has a better output power and operates at single mode. Figure 6 to Fig. 9 depict the LI curves and the optical spectra for the surface-relief VCSELs with an aperture diameter of 24 µm and 26 µm and different filter diameters. It is shown that the $0.50 \times d_{mesa}$ and $0.60 \times d_{mesa}$ wide surface-relief filters are the best options for VCSELs with apertures diameters of 24 µm and 26 µm, respectively.

## IV. SIMULATION RESULTS AND DISCUSSIONS

The LI curves of single-mode VCSELs with different aperture diameters are shown in Fig. 10. It can be seen that VCSELs with 22 µm and 26 µm apertures have the highest output powers. As expected, the VCSEL with the smaller aperture has a lower threshold current. This is because the smaller aperture causes a better carrier confinement in the active region. Therefore, the current density at the center of the 22 µm oxide aperture is higher at a certain current, which means a lower threshold current is needed. Figure 11 shows near-field patterns for single-mode VCSELs with different aperture diameters. It is shown that mode dimensions increase with increasing aperture diameter and all structures have a very narrow beam profile with the electric field intensity focused at the center.

Maximum temperature vs. injected current for single-mode surface-relief VCSELs with different aperture diameters are depicted in Fig. 12. It can be seen that as the aperture diameter increases, the temperature peak decreases. The first reason for this is that by increasing the aperture diameter, the effective resistance of the VCSEL's structure decreases. However, the temperature peaks of the VCSELs with 22 µm and 24 µm wide apertures for different injected currents are nearly similar due to the optical loss introduced by the surface-relief filter: As shown in Fig. 11, the light output power of the VCSEL with 24 µm aperture is approximately %16 lower than that of the VCSEL with an oxide aperture with a diameter of 22 µm. This means that more power is dissipated in the 24 µm VCSEL, and as a result, a higher temperature peak is observed despite having a wider oxide aperture. Overall, the VCSEL with a 26 µm wide aperture and surface-relief filter diameter of $0.60 \times d_{mesa}$ is the optimized structure. Figure 14 shows the optimal surface-relief filter diameter dependency on oxide aperture diameter.

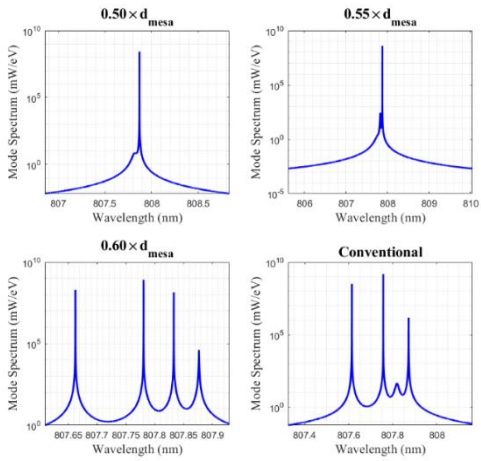

Fig. 5. Mode spectrum for different filter diameters for a VCSEL with a 22 µm wide aperture

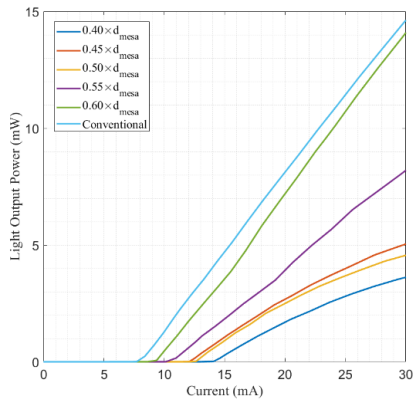

Fig. 6. LI curves for different filter diameters for a VCSEL with a 24 µm wide aperture

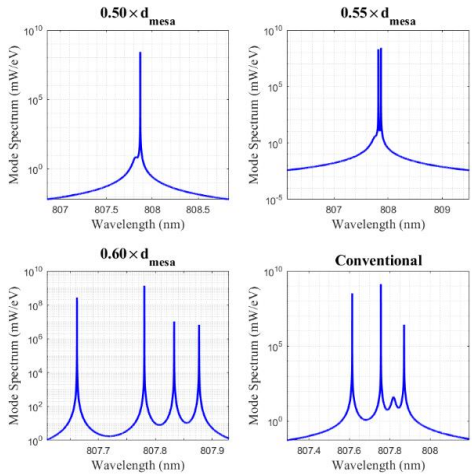

Fig. 7. Mode spectrum for different filter diameters for a VCSEL with 24 µm wide aperture

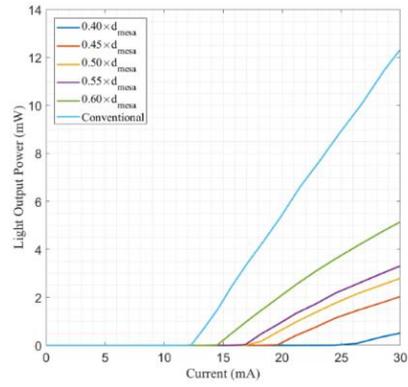

Fig. 8. LI curves for different filter diameters for a VCSEL with a 26 µm wide aperture

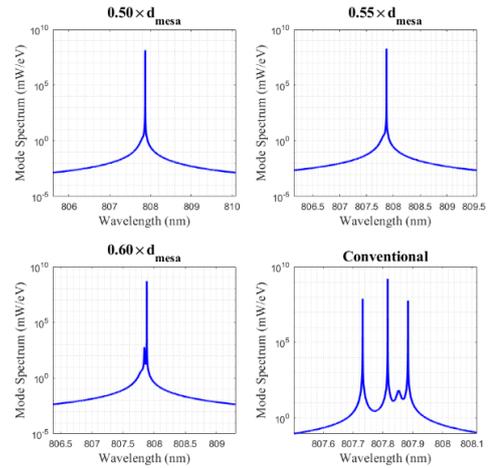

Fig. 9. Mode spectrum for different filter diameters for a VCSEL with 26 µm wide aperture

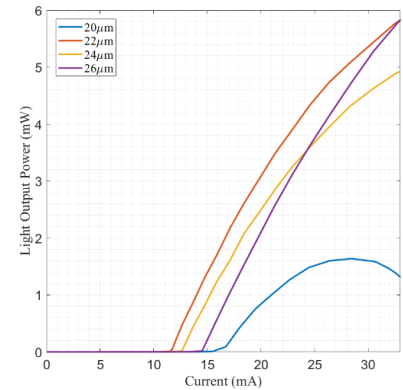

Fig. 10. LI curves for single-mode surface relief VCSELs with different aperture diameters

In order to completely understand the results of figures 10 to 13, we should consider the interrelated complexities inherent in the VCSELs. As we start to increase the oxide aperture width from 20 µm, the effective active region dimensions increase and we expect the *total* laser output power to rise. But this increase in total power is in part due to the growth of higher-order modes. As seen in Fig. 13, by increasing the aperture diameter beyond 22 µm to 25 µm, the filter diameter should be decreased to maintain single-mode operation. This means that the expected increase in the total optical power is essentially devoted to higher-order modes, forcing us to shrink filter diameters for single mode operation. This is the reason why the output power of the 24 µm VCSEL is lower than that of the 22 µm VCSEL. On the other hand, as is seen in Fig. 11, fundamental mode dimensions increase by increasing aperture dimensions (This is evident in the 26 µm VCSEL). This sets the stage for another effect to step in, which is the fact that the increased filter dimensions produce substantially higher mirror reflectivities for the fundamental mode compared to higher-order modes. Therefore, higher filter dimensions are calculated for optimum single-mode operation for apertures greater than 26 µm (See Fig. 13).

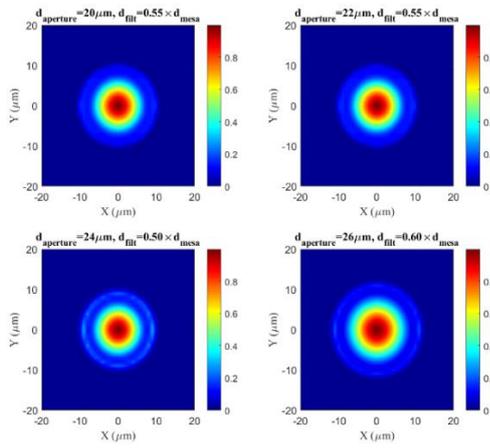

Fig. 11. Near-field patterns for single-mode surface relief VCSELs with different aperture diameters

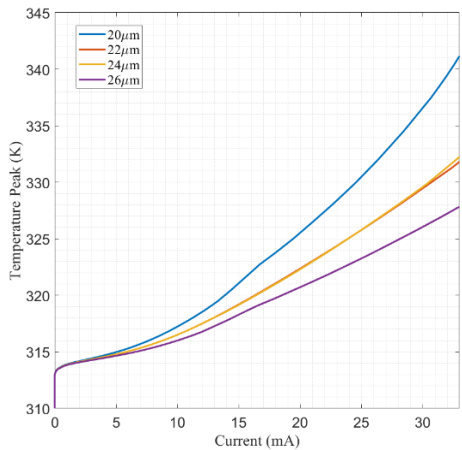

Fig. 12. Maximum temperature vs. injected current for single-mode surface relief VCSELs with different aperture diameters

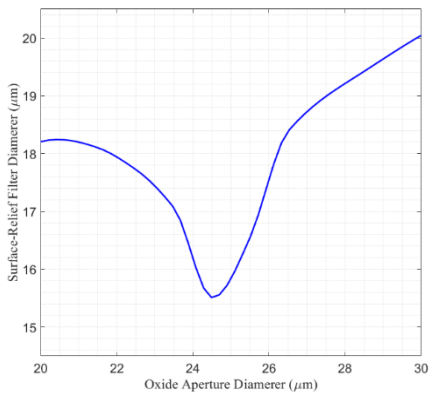

Fig. 13. Optimal surface-relief filter diameter vs. oxide aperture diameter

## V. CONCLUSION

Surface-relieves have proved to be an effective method for designing single-mode VCSELs. In this paper, we designed the optimum filter diameter for various oxide aperture sizes and studied the effect of surface-relief filter size on VCSELs' performance. We also investigated the effect of the surface-relief filter and the oxide aperture on temperature peaks in VCSELs. We realized that by increasing the diameter of the aperture, the temperature peak decreases due to a decrease in the electrical resistance of VCSEL. It is demonstrated that the VCSEL with a 26 µm wide aperture and surface-relief filter diameter of $0.60 \times d_{mesa}$ shows the highest output power and the lowest temperature peak while operating at single mode.